\documentclass[apj]{emulateapj}

\usepackage{graphicx}
\begin{document}

\title{EVLA Observations Constrain the Environment and Progenitor System of Type Ia Supernova 2011fe} 

\author{Laura~Chomiuk\altaffilmark{1}$^{,}$\altaffilmark{2}$^{,}$\altaffilmark{6}, Alicia~M.~Soderberg\altaffilmark{1}, Maxwell~Moe\altaffilmark{1}, Roger~A.~Chevalier\altaffilmark{3}, Michael~P.~Rupen\altaffilmark{2}, Carles~Badenes\altaffilmark{4}, Raffaella~Margutti\altaffilmark{1}, Claes~Fransson\altaffilmark{5}, Wen-fai~Fong\altaffilmark{1}, \& Jason~A.~Dittmann\altaffilmark{1}}
\altaffiltext{1}{Harvard-Smithsonian Center for Astrophysics, 60 Garden Street, Cambridge, MA 02138, USA}
\altaffiltext{2}{National Radio Astronomy Observatory, P.O. Box O, Socorro, NM 87801}
\altaffiltext{3}{Department of Astronomy, University of Virginia, PO Box 400325, Charlottesville, VA 22904, USA}
\altaffiltext{4}{Department of Physics and Astronomy \& Pittsburgh Particle Physics, Astrophysics, and Cosmology Center (PITT-PACC), University of Pittsburgh, Pittsburgh, PA 15260}
\altaffiltext{5}{Department of Astronomy, Stockholm University, AlbaNova, SE-106 91 Stockholm, Sweden}
\altaffiltext{6}{lchomiuk@cfa.harvard.edu}


\begin{abstract}

We report unique EVLA observations of SN\,2011fe representing the most sensitive radio study of a Type Ia supernova to date. Our data place direct constraints on the density of the surrounding medium at radii $\sim 10^{15}-10^{16}$ cm, implying an upper limit on the mass loss rate from the progenitor system of $\dot{M} \lesssim 6\times 10^{-10}~{{\rm M}_{\odot}\ {\rm yr}^{-1}}$ (assuming a wind speed of 100 km s$^{-1}$), or expansion into a uniform medium with density $n_{\rm CSM} \lesssim 6$ cm$^{-3}$. Drawing from the observed properties of non-conservative mass transfer among accreting white dwarfs, we use these limits on the density of the immediate environs to exclude a phase space of possible progenitors systems for SN\,2011fe.  We rule out a symbiotic progenitor system and also a system characterized by high accretion rate onto the white dwarf that is expected to give rise to optically-thick 
accretion winds.  Assuming that a small fraction, 1\%, of the mass accreted is lost from the progenitor system, we also eliminate much of the potential progenitor parameter space for white dwarfs hosting recurrent novae or undergoing stable nuclear burning.  Therefore, we rule out much of the parameter space associated with popular single degenerate progenitor models for SN\,2011fe, leaving a limited phase space largely inhabited by some double degenerate systems, as well as exotic single degenerates with a sufficient time delay between mass accretion and SN explosion.  

\end{abstract}

\section{Introduction}
Type Ia supernovae (SNe Ia) are luminous stellar explosions that display remarkable homogeneity in their optical properties. Comprising $\sim$30\% of all SNe in the local universe \citep{Li_etal11a}, SNe Ia are exploited as beacons for cosmography and represent the dominant source of iron-peak elements; however, we still do not understand the nature of their progenitor system(s).  There is a general consensus that SNe Ia mark the fatal disruption of white dwarfs (WDs) near the Chandrasekhar limit ($M_{\rm Ch}= 1.4~M_{\odot}$), and accretion from a binary companion is required to reach sufficient WD mass and trigger the SN explosion \citep{Hillebrandt_Niemeyer00}. However, the nature of this binary companion remains unclear, with potential progenitor systems falling into two broad classes: single degenerate (SD; containing a WD and a main-sequence, sub-giant, He star, or red-giant companion star; \citealt{Whelan_Iben73, Nomoto82}), and double degenerate (DD; in which two WDs merge; \citealt{Webbink84,Iben_Tutukov84}). In both scenarios, the evolution of the progenitor system shapes the circumbinary environment. 

Radio observations provide a sensitive probe of circumstellar material (CSM) surrounding SNe; when the shockwave plows into this material, it accelerates particles and amplifies the magnetic field, producing synchrotron emission that peaks in the cm band \citep{Chevalier82b}. The early synchrotron signal traces the CSM particle density, $n_{\rm CSM}$, on a radial scale of $\lesssim 1$ pc, the region shaped by the final stage of progenitor evolution. Radio emission is routinely detected from nearby ($d_L \lesssim 10$ Mpc) core-collapse SNe \citep{Weiler_etal02}---explosions that mark the death of massive stars, $M\gtrsim 8~M_{\odot}$ \citep{Smartt09}. However, radio emission has never been detected from a young SN Ia, despite searches spanning the past three decades \citep{Panagia_etal06, Hancock_etal11}. 

\begin{figure*}
\vspace{-1in}
\centerline{\includegraphics[width=8in]{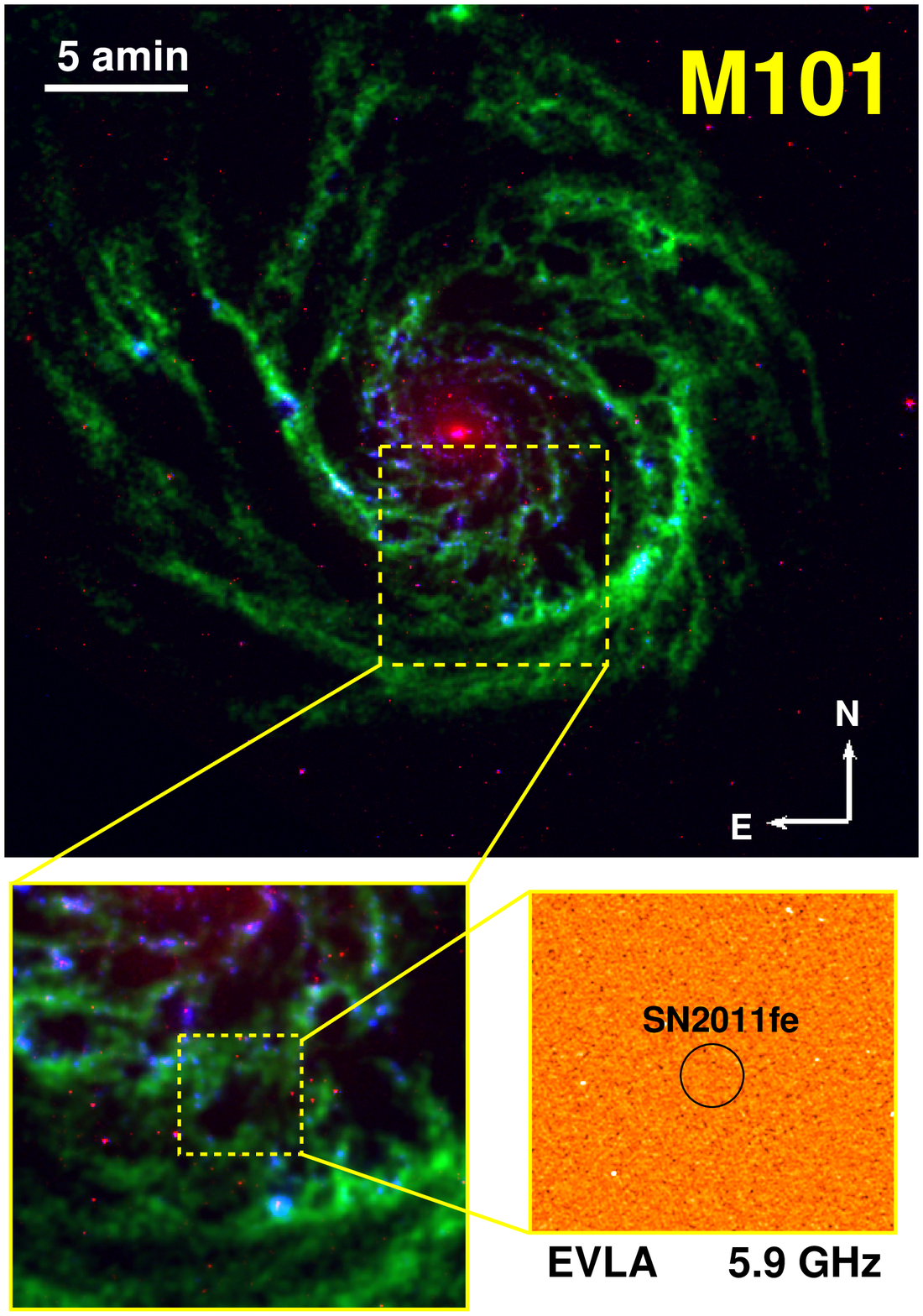}}
\vspace{-0.7in}
\caption{ Images of the environs of SN\,2011fe. The three-color image of the host galaxy, M101, features 21 cm emission (green;~\citealt{Walter_etal08}) tracing neutral hydrogen, H$\alpha$ emission (blue;~\citealt{hwb01}) tracing star formation, and 3.6 $\mu$m emission (red;~\citealt{Dale_etal09}) tracing the stellar mass. The bottom left image shows detail of the explosion site of SN\,2011fe, while the bottom right image shows our combined EVLA image at 5.9 GHz centered on the SN position (black circle has a radius of 10 arcsec).}
\label{rgb} \end{figure*}

On 2011 August 24 UT, the Palomar Transient Factory (PTF; \citealt{Law_etal09}) discovered an optical transient within the nearby spiral galaxy M101 at a distance of $6.4$ Mpc (\citealt{ss11}; Figure~\ref{rgb}).  Based on pre- and post-discovery imaging, the explosion date is well constrained to 2011 Aug 23.69 UT \citep{Nugent_etal11}.  Early optical spectroscopy revealed the transient to be a SN of the hydrogen-poor Type Ia class \citep{Nugent_etal11}, the nearest such event discovered in 25 years, dubbed SN\,2011fe (PTF11kly). Detailed optical studies indicate that SN\,2011fe is photometrically \citep{Nugent_etal11, Margutti_etal11} and spectroscopically (Parrent et al. 2012, in preparation) a normal Type Ia.

 Here we present deep Expanded Very Large Array (EVLA) radio observations of SN\,2011fe, obtained just two days after explosion and enabling a factor of ten improvement in sensitivity over previous limits.  We note that radio observations of SN\,2011fe were recently reported by \cite{Horesh_etal11}; however, our data are significantly deeper and span a longer timescale. We provide a theoretical analysis of the data that and a fresh perspective on the relation between radio luminosity and CSM density around SNe Ia. We discuss our observations and interpret the measured upper limits in Sections 2 and 3, and then consider the implications for a wide range of SN Ia progenitor scenarios and previous observations in Sections 4, 5, 6, and 7.

\section{Observations}
\begin{figure*}
\centerline{\includegraphics[width=5in, angle=90]{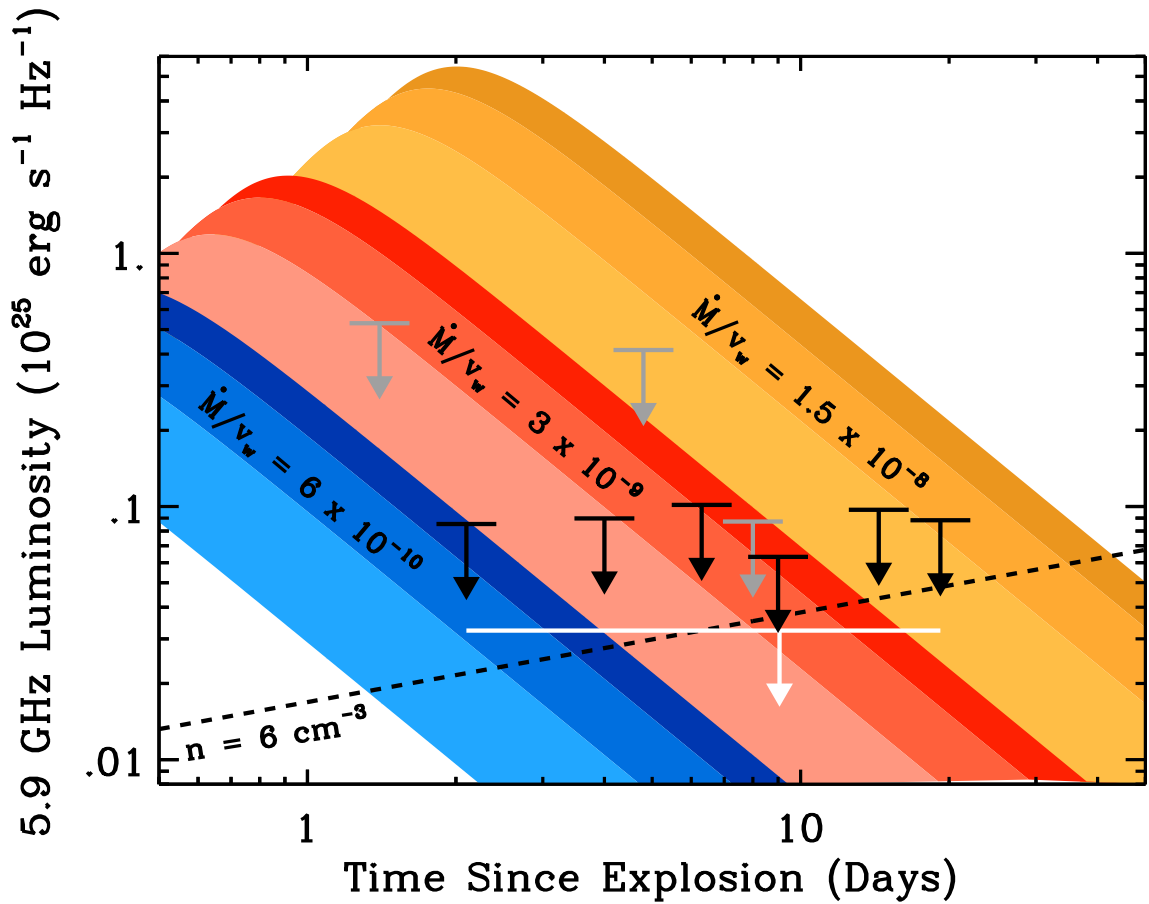}}
\caption{Deep limits on the radio luminosity of SN\,2011fe. Our 5.9 GHz upper limits (black arrows; 3$\sigma$) from six EVLA epochs are compared with model light curves for three different progenitor wind mass loss rates: $\dot{M}/v_{w} = 1.5 \times 10^{-8}\ {{{\rm M}_{\odot}\ {\rm yr}^{-1}} \over {{\rm 100\ km\ s}^{-1}}}$ (gold); $3 \times 10^{-9}\ {{{\rm M}_{\odot}\ {\rm yr}^{-1}} \over {{\rm 100\ km\ s}^{-1}}}$ (red); $6 \times 10^{-10}\ {{{\rm M}_{\odot}\ {\rm yr}^{-1}} \over {{\rm 100\ km\ s}^{-1}}}$ (blue). Each swath spans $\epsilon_B=0.01-0.1$ (with the intermediate boundaries between shades corresponding to $\epsilon_B=$0.033 and 0.066) and assumes $\epsilon_e=0.1$, $M_{\rm ej}=$ 1.4 M$_{\odot}$, and $E_{K}=10^{51}$ erg. We also plot our stacked limit as a white arrow, used to constrain a model for uniform-density CSM with a particle density of $\lesssim$ 6 cm$^{-3}$ (for $\epsilon_{B} = 0.1$; dashed black line). Finally, overplotted as grey arrows are additional data from \cite{Horesh_etal11}, scaled to 5.9 GHz.}
\label{lim} \end{figure*}

We initiated radio observations of SN\,2011fe on 2011 Aug 25.8 UT ($\Delta t\approx 2$ days after explosion) with the EVLA \citep{Perley_etal11} under our program AS1015 (PI Soderberg; \citealt{Chomiuk_Soderberg11}).  No radio counterpart was detected at the optical SN position to a limit, $F_{\nu}\lesssim 19~\rm \mu$Jy ($3\sigma$; Figure~\ref{lim}).  We continued observations of SN\,2011fe through our Director's Discretionary Time program 11B-217 (PI Soderberg) over six epochs that span three weeks following the explosion, as described in Table 1.  Our subsequent EVLA observations reveal similar upper limits (Figure \ref{lim}). Observations were carried out in A configuration using 2 GHz of bandwidth and recording four polarization products; one baseband of 1 GHz width was centered at 5.0 GHz while the other was centered at 6.75 GHz. Each epoch had one hour duration, yielding 40 minutes on source, except 2011 Aug 26, when we observed for two hours. Data were calibrated using J1349+5341 and 3C286, and were reduced using standard routines in the Astronomical Image Processing System (AIPS). Each sideband was imaged individually; the 6.75 GHz image was then smoothed to the resolution of the 5.0 GHz image, and the two noise-weighted images were averaged together.
We also created a stacked image by concatenating all {\it uv} data for a given sideband and producing a final image from the combined data set. Again, the two sidebands were averaged together in the image plane, producing an image that reaches the thermal noise limit at 5.9 GHz. Our stacked EVLA data place a strong constraint on the radio luminosity of $L_{\nu} \lesssim 3.2\times 10^{23}~\rm erg~s^{-1}~Hz^{-1}$ on an averaged date of 9.1 days after explosion (Figure~\ref{rgb}).

The resolution of our EVLA images is $0.56^{\prime\prime} \times 0.39^{\prime\prime}$. The flux densities quoted in Table 1 are measured at the SN coordinates given by \cite{Li_etal11b}. The nearest point displaying any significant flux in our stacked image is a 4.5$\sigma$ (10.2 $\mu$Jy at maximum) point-like peak located 0.7$^{\prime\prime}$ away from the \cite{Li_etal11b} position (at J2000.0 RA = $14^{\rm h}03^{\rm m}05.799^{\rm s}$, Dec = $+54^{\circ}16^{\prime}24.75^{\prime\prime}$). However, when we image the RR and LL correlations of the EVLA polarizations separately, we find that no source is visible in the RR image ($3 \pm 3\ \mu$Jy in RR, as opposed to $16 \pm 3\ \mu$Jy in LL), thus implying that it is it is likely an instrumental artifact, and unlikely to be a real source. We also investigated if the position of the radio peak could possibly be consistent with SN\,2011fe by checking the world coordinate systems of our EVLA image and the {\it HST} ACS/F814W image (GO-9490, PI K.~Kuntz) used by \cite{Li_etal11b} for consistency. We cross-compared the positions of two compact \ion{H}{2} regions located $\sim$1.5$^{\prime}$ northeast of SN\,2011fe that are visible in both the \emph{HST} and EVLA images. For this comparison, we also made use of a narrow-band H$\alpha$ image from {\it HST}/WFPC2 (GO-5210, PI J.~Trauger), as these \ion{H}{2} regions are much more clearly defined in F656N than in F814W. We find that the coordinate systems are not offset by more than 0.2$^{\prime\prime}$, and thus the 4.5$\sigma$ radio peak is positionally inconsistent with SN\,2011fe. 

\begin{deluxetable}{lccc}
\tablewidth{0 pt}
\tablecaption{ \label{tab:obs}
 EVLA Observational Data for SN\,2011fe}
\tablehead{Date & Flux Density & Time Since Explosion \\
(UT) & ($\mu$Jy) & (Days) }
\startdata
2011 Aug 25.77 & $7.8\pm5.8$ & 2.1 \\
2011 Aug 27.72 & $0.4\pm6.1$ & 4.0\\
2011 Aug 29.98 & $-8.4\pm6.9$ & 6.3\\
2011 Sept 2.68 & $0.7\pm4.3$ & 9.0 \\
2011 Sept 8.04 & $-2.7\pm6.6$ & 14.4 \\
2011 Sept 12.93 & $0.6 \pm6.0$ & 19.2\\
\hline
Stacked: Aug 25.8--Sept 12.9 & $-0.6\pm2.2$ & 2.1--19.2\\
\enddata
\end{deluxetable}

\section{Density Constraints from Radio Upper Limits}

We make predictions for the radio emission from SNe Ia by analogizing them to other hydrogen-poor SNe which are detected in the radio, namely Type Ib and Type Ic SNe (SNe Ibc; e.g., \citealt{Berger_etal03, cf06, Soderberg_etal10}). SNe Ibc are generally understood to mark the gravitational core collapse of massive stars that have been stripped of their hydrogen envelopes prior to explosion.   The analogy between SNe Ia and Ibc is motivated by several commonalities, including (i) their compact progenitors, leading to mildly-relativistic shockwaves at the time of breakout from the star \citep{Colgate70, Nakar_Sari11}, and (ii) a common polytrope of $\gamma = 4/3$, appropriate for both massive (relativistic) WDs and the radiative envelopes of Wolf-Rayet stars.  It is estimated that for SNe arising from compact progenitors, the outermost density profile of the ejecta (hosting the fastest moving material, which gives rise to the radio emission) is characterized by a steep power-law,  $\rho_{\rm SN} \propto r^{-10.2}$; \citep{mm99}. The coupling of mass to ejecta velocity is simply a function of the bulk explosion energy ($E_{51}$, in units of $10^{51}$ erg) and ejecta mass ($M_{\rm ej}$; \citealt{Berger_etal02, cf06}).  These parameters are well constrained in the case of SNe Ia; $E_{51} = 1$ and $M_{\rm ej}=M_{\rm ch} = 1.4~M_{\odot}$ are fiducial values \citep{Mazzali_etal07}.  

The interaction of the shockwave with the immediate environment is then described by a self-similar solution, such that the evolution of the shockwave radius ($R_s$) and its velocity ($v_{s}$) are determined by the properties of the environment: its particle density ($n_{\rm CSM}$) and its radial profile \citep{Chevalier82}.   This dynamical interaction accelerates particles from the circumstellar environment into a power-law energy distribution, $N(E) = N_{0}\ E^{-p}$, where $N(E)$ is the density of relativistic electrons at a given energy $E$, above a minimum energy, $E_m$.  We further assume that constant fractions of the post-shock energy density is shared between relativistic electrons ($\epsilon_e$) and amplified magnetic fields ($\epsilon_B$) and a shock compression factor, $\eta=4$ \citep{cf06}.
In this framework,
\begin{equation}
B = \sqrt{8 \pi\ \epsilon_B\ \rho_{\rm CSM}\ v_{s}^2},
\end{equation}
\begin{equation}
N_{0} = (p-2)\ \epsilon_e\ \rho_{\rm CSM}\ v_{s}^2\ E_{m}^{p-2},
\end{equation}
and
\begin{equation}
E_m = {{(p-2) \epsilon_e \mu m_{p} v_{s}^2} \over {(p-1) \eta (n_e/n_i)}}.
\end{equation}
Here $\mu$ is the mean molecular weight of the CSM assuming it is neutral, and ($n_e/n_i$) is the number ratio of electrons to ions in the shocked gas (for solar abundance, $\mu=1.4$ and $n_e/n_i = 1.14$). The shock-accelerated electrons gyrate along the magnetic field lines, giving rise to synchrotron radiation that peaks in the radio shortly after explosion with a spectrum, $L_{\nu}\propto \nu^{\beta}$; here $\beta=-(p-1)/2$.  The high velocity shockwaves of Type I SNe minimize opacity to external free-free absorption, such that synchrotron self-absorption dominates the spectral energy distribution, suppressing the emission at low frequencies \citep{c98}. For sources characterized by synchrotron self absorption, the shock radius may be robustly inferred from the synchrotron light curve
\citep{Readhead94}, thus enabling a direct tie between the observed radio properties and the SN self-similar solution \citep{c98}.

For SN\,2011fe, we assume $p=3$,  $\epsilon_e=0.1$, and a range of $\epsilon_B=[0.01-0.1]$ as inferred for SN Ib/c \citep{cf06,Zhang_etal09,Soderberg_etal11}.  In this framework, the radio luminosity at a given frequency and time after explosion depends only on $n_{\rm CSM}$, and is calculated using Equation 1 of \cite{c98}
Therefore, radio observations probe the environment of SNe Ia and may distinguish between different progenitor systems (see Chomiuk et al. 2012, in preparation, for a detailed discussion).

\subsection{Wind Density Profile for the CSM}
A commonality among SD progenitor channels is the accretion of material onto a WD due to its interaction with a donor star. While possible progenitor scenarios span a wide parameter space \citep{Hillebrandt_Niemeyer00}, most scenarios may be characterized by the WD's accretion rate, $\dot{M}_{\rm acc}$, and efficiency of accretion, $\epsilon_{\rm acc}$, such that the rate of mass lost into the local environs is $\dot{M}_{loss} = (1-\epsilon_{\rm acc}) \dot{M}_{\rm acc}$ \citep{Nomoto_etal07}.  Here we define, $\epsilon_{\rm loss}=(1-\epsilon_{\rm acc}$).  In this non-conservative mass transfer scenario, we assume that the lost material is blown outward with a wind velocity, $v_w$, and radially distributed as $n_{\rm CSM} = \dot{M} / (4 \pi v_{w} r^{2})$. 

Using the similarity solutions of \cite{Chevalier82} and the SN Ia ejecta density profile of Chomiuk et al.~(2012, in preparation), we estimate the shockwave radius to be:
\begin{equation}\label{eq:r_wind}
R_{s} = 5.5 \times 10^{14}\ (\dot{M}/v_{w})^{-0.12}\ E_{51}^{0.43}\ M_{\rm ch}^{-0.31}\ t_{10}^{0.88}\ {\rm cm},
\end{equation}
where $t_{10}$ is the time since explosion in units of 10 days. For $t_{10}=0.21, \dot{M}/v_{w} = 10^{-10} {{{\rm M}_{\odot}\ {\rm yr}^{-1}} \over {{\rm km\ s}^{-1}}}, E_{51} = 1$, and $M_{\rm ch} =1$, we find $R_{s} = 2.2 \times 10^{15}$ cm and the shockwave velocity ($v_{s} = 0.88 R_{s}/t$) is $0.35 c$. 

When the SN shockwave interacts with such a medium, the radio emission is suppressed at early times by synchrotron self absorption, and declines at later times due to the decelerating shockwave and decreasing density of the CSM.  With the radius evolution and assumptions described above, we calculate a family of radio light curves using the formalism of \cite{c98} and \cite{cf06}. In the optically-thin regime (after light-curve peak, which is valid here; Figure \ref{lim}), for a given luminosity limit, our constraint on $\dot{M}/v_w$ scales as $\epsilon_{B}^{-0.7}$.

Our measured upper limits on $L_{\nu}$ measured $\Delta t\approx 2.1$ days after explosion provide the deepest limits on the density of the environment surrounding SN\,2011fe to date, corresponding to $\dot{M}/v_w\lesssim 6\times 10^{-10}~{{{\rm M}_{\odot}\ {\rm yr}^{-1}} \over {{\rm 100\ km\ s}^{-1}}}$ (assuming $\epsilon_B = 0.1$) or $\dot{M}/v_w\lesssim 3\times 10^{-9}~{{{\rm M}_{\odot}\ {\rm yr}^{-1}} \over {{\rm 100\ km\ s}^{-1}}}$ (assuming $\epsilon_B = 0.01$) for the SN\,2011fe progenitor system. These limits are a factor of $\sim 20$ deeper than those presented by \cite{Horesh_etal11}, due to our calculation from first principles of $v_s$ and our self-consistent treatment of $E_m$.  

Mass loss from a SD progenitor may be concentrated toward the equatorial or polar regions of the orbit \citep[e.g.,][]{Mohamed_Podsiadlowski11}. Assuming that the asymmetric mass loss still follows an $n_{\rm CSM} \propto r^{-2}$ radial profile, and covers a solid angle $\Omega$ (units of sterradians), the radio luminosity scales as $L_{\nu} \propto \left({{\Omega}\over{4 \pi}}\right)^{-0.4}$. The increase in CSM density due to asymmetry wins out over the slightly stronger deceleration of the shockwave, leading to an increase in the radio luminosity for a given $\dot{M}/v_w$, if material is expelled asymmetrically. Therefore, our assumption of spherical symmetry is conservative.

\subsection{Uniform Density Profile for the CSM}
We also consider a uniform density medium, as might be expected if the SN is exploding into the ambient interstellar medium. In this case, again using the similarity solutions from \cite{Chevalier82} and the density profile of SN Ia ejecta from Chomiuk et al.~(2012, in preparation), the SN shockwave radius evolves as:
\begin{equation}
R_{s} = 1.4 \times 10^{16}\ n_{\rm CSM}^{-0.10}\ E_{51}^{0.35}\ M_{\rm ch}^{-0.25}\ t_{10}^{0.71}\ {\rm cm}.
\end{equation}
For $t_{10}=0.21, n_{\rm CSM} = 1\ {\rm cm}^{-3}, E_{51} = 1$, and $M_{\rm ch} =1$, we calculate $R_{s} = 4.5 \times 10^{15}$ cm and a shockwave velocity ($v_{s} = 0.71 R_{s}/t$) of $0.58 c$. The radio light curve in the case of a uniform density medium is relatively flat, as compared the wind density profile (Figure~\ref{lim}), so the most powerful observational constraint comes from our stacked limit, giving a density limits in the range $n_{\rm CSM} \lesssim 6$ cm$^{-3}$ ($\epsilon_B = 0.1$) to $n_{\rm CSM} \lesssim 44$ cm$^{-3}$ ($\epsilon_B = 0.01$). Here, our limits on the density scale with the assumed value for $\epsilon_{B}$ as $n_{\rm CSM} \propto \epsilon_B^{-0.9}$. It is noteworthy that, unlike in the wind case, the radio luminosity actually increases with time, because the decline in post-shock energy density is more gradual. In this uniform density case, the growth in the radio-emitting volume overwhelms the effects of gradual deceleration of the shockwave, implying that the total energies in magnetic fields and relativistic electrons slowly grow as the SN expands. 

\section{Implications for SN Ia Progenitors}

\begin{figure*}
\centerline{\includegraphics[width=7in]{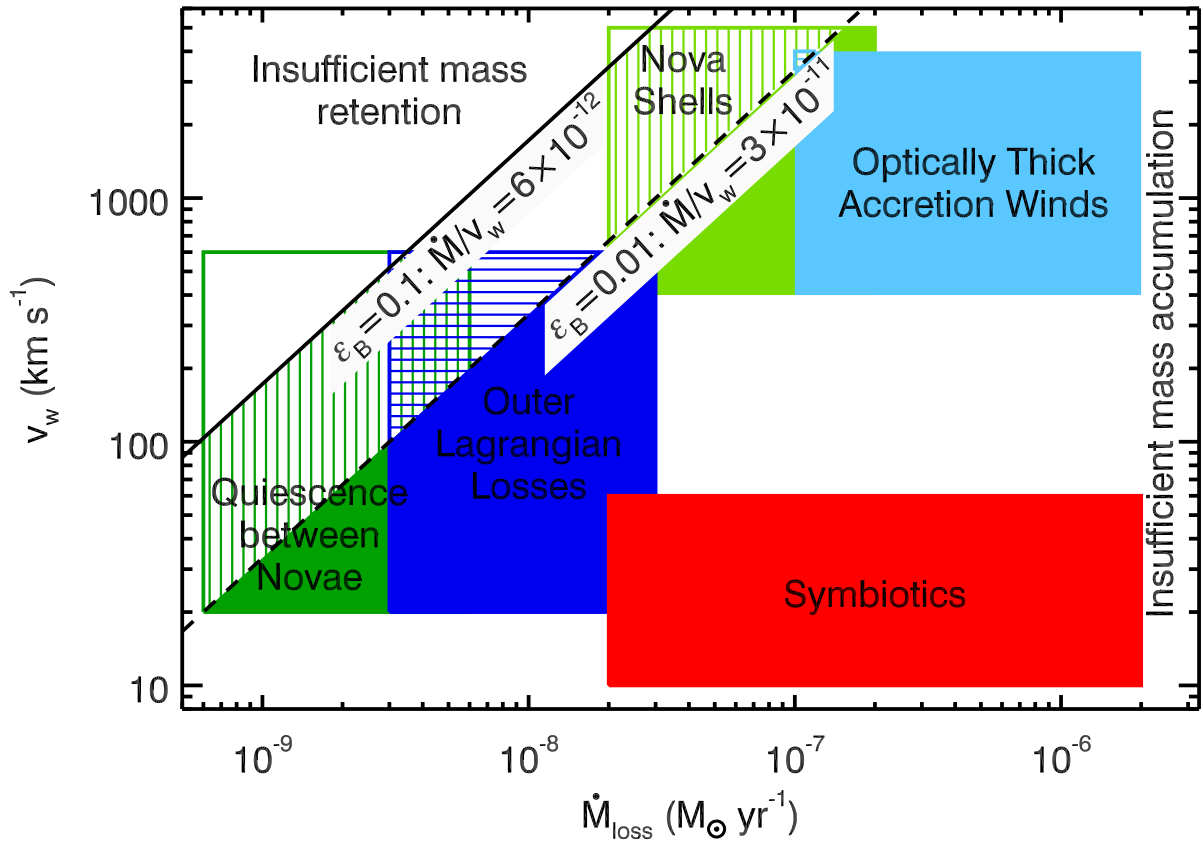}}
\caption{Constraints on the parameter space of mass loss rate from the binary system versus the velocity of the lost material. 
The progenitor scenarios discussed in Section 4 are plotted as schematic zones in $\dot{M}$--$v_{w}$ space. We indicate our 3$\sigma$ limits on $\dot{M}/v_{w}$ assuming $\epsilon_B$ = 0.1 (solid; corresponding to the dark blue curve in Figure~2), and the conservative case of $\epsilon_B$ = 0.01 (dashed; corresponding to the light red curve in Figure~2).  Mass loss scenarios to the lower right of these lines would have been detected with our EVLA observations, and are eliminated for SN\,2011fe. 
At the lowest mass accretion rates, nova eruptions will expel the accreted material, and there will be insufficient mass retention toward SNe Ia. At high mass transfer rates $\gtrsim 2 \times 10^{-6}\ {{\rm M}_{\odot}\ {\rm yr}^{-1}}$, too much mass is lost though winds, so that the WD will not accrete sufficient mass to approach $M_{\rm Ch}$.}
\label{mdotv} \end{figure*}

Below, we compare these constraints on the environment of SN\,2011fe with predictions for popular single degenerate scenarios, summarized in Figure \ref{mdotv}.

\noindent \emph{\textbf{Symbiotic Systems:}} In the first SD scenario, the SN\,2011fe progenitor system consists of a WD in a symbiotic binary system, accreting mass from a giant star (Figure~\ref{mdotv}). Mass is lost from the system at a rate $\dot{M} > 10^{-8}~M_{\odot}~\rm yr^{-1}$ with velocity $v_w\approx 30 ~\rm km~s^{-1}$ \citep{Seaquist_Taylor90,Patat_etal11a,Chen_etal11}. This model is ruled out by our EVLA observations and is consistent with the non-detection of a giant star in pre-explosion optical imaging \citep{Li_etal11b} and X-ray limits on the inverse Compton emission from the shockwave (\citealt{Horesh_etal11}, \citealt{Margutti_etal11}).

\noindent \emph{\textbf{White Dwarfs with Steady Nuclear Burning:}} Next, we consider a main sequence, subgiant, or helium star undergoing stable Roche Lobe overflow onto a WD. For accretion rates of  $\dot{M}_{\rm acc} \gtrsim 3\times 10^{-7}~\rm M_{\odot}~yr^{-1}$, the WD will undergo steady nuclear burning \citep{Shen_Bildsten07}.  We assume that a small fraction of the transferred mass ($\epsilon_{\rm loss} \approx$1\%) is lost at the outer Lagrangian points of the system and travels outwards at at some factor of order unity of the orbital velocity, $v_{w}\approx 100~\rm km~s^{-1}$ (dark blue region in Figure \ref{mdotv}). This mode of outer Lagragian mass loss with velocities up to $\sim$600 km s$^{-1}$ has been manifested in the P-Cygni profiles of stable nuclear burning WDs undergoing these moderate accretion rates \citep{Deufel_etal99}. This scenario is also supported by the circumbinary material observed from cataclysmic variable systems \citep{Williams_etal08}, as well as the steady orbital separations measured in long period cataclysmic variables \citep{Huang_Yu96}. Most of this parameter space is ruled out for SN\,2011fe by our EVLA limits, in the $\epsilon_B = 0.1$ case.

At higher mass transfer rates, optically-thick winds in the vicinity of the WD are hypothesized to limit the mass accretion to $\dot{M}_{\rm acc}\approx 6\times 10^{-7}~\rm M_{\rm \odot}~yr^{-1}$, and any additional mass transferred is lost from the system with velocity, $v_{w} \approx$ few $\times$ 1,000 km s$^{-1}$ \citep[light blue region in Figure \ref{mdotv};][]{Hachisu_etal99}, consistent with outflows seen in the most X-ray luminous nuclear burning WDs \citep{Cowley_etal98}. This scenario is completely excluded by our EVLA limits if the accretion wind immediately preceded the SN explosion (see \S 6.1 for further discussion). 

Near the transition point between these moderate and high accretion rate regimes, spectroscopic analysis of the absorption profiles of luminous nuclear-burning WDs reveals that the mass loss rate is $\sim$10\% the accretion rate, or $3 \times 10^{-8}~M_{\odot}~\rm yr^{-1}$, and is expelled at a high velocity of $v_{w}$ = 3,000 km s$^{-1}$ \citep{Murray02}. Such a high-velocity accretion wind would have been barely detectable by our radio data (assuming $\epsilon_{B} = 0.1$, see Fig. 3); nevertheless, the slower wind via the outer Lagragian points, albeit intrinsically lower in its mass loss rate, would have been readily detectable due to the high accretion rate in such a system. 

Our EVLA observations of SN\,2011fe therefore critically constrain stable nuclear burning in both accretion regimes (outer Lagrangian losses and optically-thick winds; Figure~\ref{mdotv}).  Parallel studies based on pre-explosion X-ray images of the SN\,2011fe progenitor system only rule out a subset of these systems with the most compact and hottest photospheres \citep{Li_etal11b, Liu_etal11}.

\noindent \emph{\textbf{Recurrent Novae:}} Finally, at lower accretion rates, $\dot{M}_{\rm acc}\approx (1-3) \times 10^{-7}$ M$_{\odot}$ yr$^{-1}$, the WD may undergo nova outbursts with a short recurrence time of several years. The nova shells expand with velocities of $1,000-5,000$ km s$^{-1}$, leaving evacuated cavities between the shells \citep{Wood-Vasey_Sokoloski06}. In the paradigm of synchrotron emission produced by the interaction of SN shock with CSM \citep{Chevalier82b}, radio emission will be generated when the SN shockwave crashes through a previously ejected shell. 

\begin{figure*}
\centerline{\includegraphics[width=5in, angle=90]{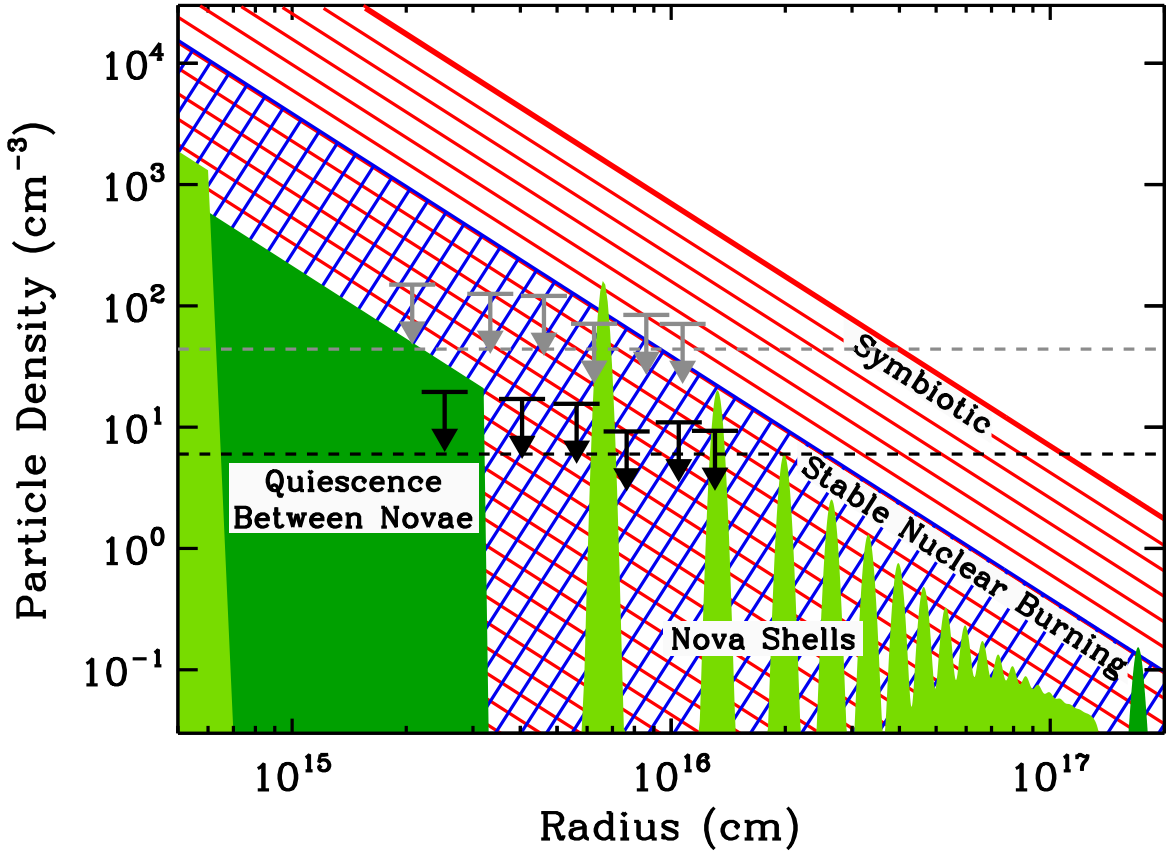}}
\caption{Theoretical CSM radial density profiles are estimated for four possible SD progenitor scenarios. {\it Red:} a symbiotic progenitor system with $\dot{M} = 1 \times 10^{-7}$ M$_{\odot}$ yr$^{-1}$ and $v_{w}$ = 30 km s$^{-1}$; {\it Blue:} stable nuclear burning, for either outer Lagrangian losses ($\dot{M} = 2 \times 10^{-8}$ M$_{\odot}$ yr$^{-1}$ and $v_{w}$ = 100 km s$^{-1}$) or optically-thick winds ($\dot{M} = 2 \times 10^{-7}$ M$_{\odot}$ yr$^{-1}$ and $v_{w}$ = 1,000 km s$^{-1}$); and {\it Green:} a recurrent nova system with short recurrence timescale (2 years; light) and longer recurrence timescale (10 years; dark).  Assuming a $n_{\rm CSM} \propto r^{-2}$ CSM density profile, we estimate the CSM densities and shockwave radii that would correspond to our 5.9 GHz limits, for $\epsilon_B=0.1$ (black arrows) and $\epsilon_{B}=0.01$ (grey arrows). Also overplotted are limits for uniform density surroundings, again for $\epsilon_B=0.1$ (black dashed line) and $\epsilon_{B}=0.01$ (grey dashed line).}
\label{dprof} \end{figure*}

For SN\,2011fe, our EVLA observations probe a circumbinary scale of $r\approx 10^{16}$ cm, thus constraining the presence of shells from novae with short recurrence time, $\sim$few years. The light green zones in Figure \ref{dprof} show the distribution of CSM around a nova with recurrence time of 2 years. Models of such a system show that $2 \times 10^{-7}$ M$_{\odot}$ yr$^{-1}$ is incident upon the WD \citep{Yaron_etal05,Shen_Bildsten09}; we assume that 1\% of this transferred material is lost to the binary to form the CSM at small radii. During a nova explosion, for the adopted accretion rate onto a $\sim 1.4$ M$_{\odot}$ WD, models imply that $\sim$15\% of the material accreted over the last two years is ejected in a shell with a mass of $6 \times 10^{-8}$ M$_{\odot}$ (light green region in Figure \ref{mdotv}). Adopting a shell thickness of $\sim 0.1 R$, the radio emission is likely short-lived ($\sim$days) and we estimate a $\sim$30\% probability of radio detection of such nova shells in our EVLA data.  

For longer recurrence times, the progenitor wind may enhance the local CSM density between eruptions (e.g., U Sco; recurrence timescale of $\sim$10 years). The dark green regions plotted in Figure \ref{dprof} represent a nova with recurrence time of 10.3 years, modeled after U Sco. In the case of U Sco, a nova ejecta mass of $10^{-6}$ M$_{\odot}$ is expelled every $\sim$10 years at a velocity of 5,000 km s$^{-1}$ \citep{Hachisu_etal00, Schaefer10, Diaz_etal10}. Between eruptions, mass is transferred to the WD at a rate $\dot{M}_{\rm acc} = 10^{-7}$ M$_{\odot}$ yr$^{-1}$ \citep{Duschl_etal90,Hachisu_etal00}. Again, we assume $\epsilon_{\rm loss} \approx$ 1\% of this transferred mass is lost through the outer Lagrangian points in a wind expelled at roughly the orbital velocity, $v_{w} = 100$ km s$^{-1}$ (dark green region in Figure \ref{mdotv}). The local density is therefore $\dot{M}/v_{w} = 1\times 10^{-9} {{\epsilon_{\rm loss}} \over {0.01}} {{{\rm M}_{\odot}\ {\rm yr}^{-1}} \over {{\rm 100\ km\ s}^{-1}}}$, and comparable to our EVLA detection limits (Figure~\ref{dprof}) but undetectable in pre-explosion optical images \citep{Li_etal11b}. 

Our assumed ejected and accreted masses for U Sco are intermediate in the uncertain range of estimates in the literature; for example, measurements of the ejecta mass could be a factor of three higher or lower \citep{Evans_etal01,Diaz_etal10}, and estimates of the accretion rate are model dependent. These uncertainties underline our point that we can not conclusively rule out a U Sco-like progenitor for SN\,2011fe, although we can constrain the large potential parameter space for U Sco-like progenitors.

\section{Estimates of the Broader Environment of SN\,2011fe}

Alternatively, if SN\,2011fe is the result of a DD progenitor system, it may have exploded into an undisturbed and constant density environment. As shown in Figure~\ref{rgb}, we can probe the ambient interestellar medium in the $\sim$0.3 kpc region surrounding the SN explosion site using pre-explosion \ion{H}{1} 21 cm imaging from The \ion{H}{1} Neary Galaxy Survey \citep[THINGS;][]{Walter_etal08}. We measure an integrated brightness temperature of 165.4 K km s$^{-1}$ at the location of SN\,2011fe, corresponding to a \ion{H}{1} column density of $3.02 \times 10^{20}$ cm$^{-2}$. Assuming a path length of 100 pc through the face-on neutral disk of M101 and solar abundance, we roughly estimate a volume number density of $\sim$1 cm$^{-3}$ at this location, typical of the warm phase of the interstellar medium \citep{Ferriere01}. Our stacked EVLA limits, which imply $n_{\rm CSM} \lesssim (6-44)$ cm$^{-3}$, are therefore consistent with expansion into the ambient ISM. 

Using H$\alpha$ imaging as a tracer of star formation in M101, we can also compare the location of SN\,2011fe relative to the host galaxy's light distribution. We adopt the methodology for calculating a fractional flux value from \cite{Fruchter_etal06} and \cite{Kelly_etal08}. We obtain an H$\alpha$ narrow-band image of the field from \cite{hwb01} with a central wavelength of 6564 \AA, and calculate the fraction of total host light in pixels fainter than the SN position. To determine an appropriate cut-off level for the host, we create an intensity histogram of a $30' \times 30'$ region centered on M101 and model the sky brightness distribution with a Gaussian profile, taking pixels with flux above $3\sigma$ to be part of M101. We then calculate the flux in galaxy pixels fainter than the location of SN\,2011fe and normalize by the total galaxy flux. This gives a fractional flux value of $0.55$, showing that SN\,2011fe originates from an ordinary region in a star-forming galaxy, consistent with results for other SNe Ia \citep{Kelly_etal08}. This result affirms that SN\,2011fe exploded in a typical environment for a SN Ia, and is not closely associated with a young stellar population (in contrast, SNe Ic and long gamma-ray bursts, which are thought to closely follow star formation, typically have significantly higher fractional fluxes than SNe Ia). 

\section{Discussion}
\subsection{Constraints on the Environments of SNe Ia}
Here we consider our characterization of the SN\,2011fe environment in the context of other multi-wavelength studies of SN Ia environs.

The interaction of a SN shockwave with surrounding CSM should also produce X-ray and H$\alpha$ emission, along with the radio emission described in detail here. To date, searches for this emission from SNe Ia have yielded only non-detections, although they typically place weaker constraints on the CSM density than radio limits (however, \citealt{Margutti_etal11} have recently shown that deep X-ray observations obtained at optical peak have the potential to surpass radio constraints). Early-time deep H$\alpha$ observations of SN 2001el constrain the mass loss rate of the progenitor to $\dot{M}/v_{w} < 10^{-4}\ {{{\rm M}_{\odot}\ {\rm yr}^{-1}} \over {{\rm 100\ km\ s}^{-1}}}$ \citep{Mattila_etal05}. For the normal SN Ia 2002bo, \cite{Hughes_etal07} used X-ray observations to constrain $\dot{M}/v_{w} < 2 \times 10^{-4}\ {{{\rm M}_{\odot}\ {\rm yr}^{-1}} \over {{\rm 100\ km\ s}^{-1}}}$, assuming thermal emission dominates the X-ray signal. They also revisit a previously claimed X-ray detection of SN\,2005ke \citep{Immler_etal06}, showing that this source is in fact not present in more carefully analyzed X-ray data. Recently, all \emph{Swift} X-ray observations of SNe Ia to date have been stacked by \cite{Russell_Immler12}, yielding a non-detection of $L_{X} < 1.7 \times 10^{38}$ erg s$^{-1}$ in the energy range 0.2--10 keV. Radio limits on CSM around SN\,2011fe are consistent with deep \emph{Chandra} observations obtained 4 days after explosion, which place an upper limit $\dot{M}/v_{w} < 2 \times 10^{-9}\ {{{\rm M}_{\odot}\ {\rm yr}^{-1}} \over {{\rm 100\ km\ s}^{-1}}}$ (\citealt{Margutti_etal11}; see also \citealt{Horesh_etal11}).

Our EVLA upper limits are a strong indication that there is not much CSM within $\sim 10^{16}$ cm of the progenitor of SN\,2011fe, consistent with a uniform density medium of $n_{\rm CSM} \lesssim$ 10 cm$^{-3}$. Similar constraints can be drawn from the morphology and X-ray emission of historical supernova remnants (SNRs) of Type Ia origin.  For example, X-ray studies of SNR\,0509-67.5 in the Large Magellanic Cloud and the Tycho SNR (both $\sim$400 years old) imply that these SNRs are surrounded by uniform media of density $\sim$0.5--1 cm$^{-3}$ \citep{Badenes_etal06, Badenes_etal08}, typical of the warm phase of the interstellar medium. H$\alpha$ imaging of SN\,1006 with \emph{HST} shows that this remnant is expanding into a medium of density 0.25--0.4 cm$^{-3}$, and the medium is relatively uniform, with fluctuations of $\sim$20\% on length scales of $\sim$1 pc \citep{Raymond_etal07}. The youngest SNR in the Milky Way, G1.9+0.3, is likely the remnant of a SN Ia that exploded $\sim$100 years ago, and is also consistent with low-density surroundings \citep[$\sim$0.03 cm$^{-3}$;][]{Reynolds_etal08, Borkowski_etal10, Carlton_etal11}. This widespread consistency with our results is remarkable, as these SNRs have radii of $\sim 10^{19}$ cm, three orders of magnitude larger than the extent probed by our EVLA data. 

SNRs can also shed additional light on the optically-thick accretion wind scenario, postulated by \cite{Hachisu_etal99} to blow from WDs that are accreting near the Eddington rate ($\gtrsim 6 \times 10^{-7}$ M$_{\odot}$ yr$^{-1}$). We have already shown that our radio observations can exclude such a wind in SN\,2011fe if it immediately precedes the SN explosion, but in other evolutionary scenarios there may be a delay between the production of the wind and the SN of $\sim10^{5}-10^{6}$ years \citep{Han_Podsiadlowski04}. In this case, a large cavity will surround the WD at the time of SN Ia explosion, with a radius of $\sim$10 pc and $n_{\rm CSM} \lesssim 10^{-3}$ cm$^{-3}$ \citep{Badenes_etal07}, and the outer edge of this cavity will be marked by a dense shell with $n_{\rm CSM} \approx 100$ cm$^{-3}$. This scenario would yield radio non-detections at early times, as we observe, but would also leave a clear imprint on SNRs, wherein SNRs should have larger radii and faster expansion velocities than expected for typical interstellar medium densities. Such an effect is  not observed in a sample of seven Type Ia SNRs \citep{Badenes_etal07}; combined with our early time observations, this implies that accretion winds can not play a major role in the progenitor systems of most SNe Ia. An interaction with the warm phase of the interstellar medium seems to be able to explain the properties of most Type Ia SNRs.

Another intriguing constraint on the CSM surrounding SNe Ia comes from high-resolution optical spectroscopy, as observations of time-variable narrow \ion{Na}{1} D absorption lines in the spectra of SNe Ia \citep{Patat_etal07, Blondin_etal09, Simon_etal09, Sternberg_etal11}. Assuming that the variations in time are due to ionization by the SN light followed by gradual recombination, the distance to the absorbing material can be estimated at $\gtrsim 10^{17}$ cm with a mass of $10^{-5}-10^{-2}$ M$_{\odot}$, depending on its geometry and clumpiness \citep{Chugai08, Simon_etal09}. This distribution of material may be consistent with old nova shells from a recurrent nova (similar absorption features were observed following the 2006 outburst of RS Oph by \citealt{Patat_etal11a}. However, U Sco did not show time-variable absorption lines during its 2010 outburst; \citealt{Kafka_Williams11}). We note that in the case of SN\,2006X (the first published SN Ia with recognized time-variable \ion{Na}{1} D lines), VLA observations were acquired and yielded a non-detection, constraining $\dot{M}/v_w\lesssim 3\times 10^{-7}~{{{\rm M}_{\odot}\ {\rm yr}^{-1}} \over {{\rm 100\ km\ s}^{-1}}}$ at a distance of $\sim 10^{17}$ cm \citep[][using the models described here]{Patat_etal07}.

As we have shown, early-time radio observations can only rule out some of the parameter space associated with recurrent novae, and the observations presented here are sensitive to material at radii $\lesssim 10^{16}$ cm, so high-resolution optical spectroscopy is a complementary strategy for probing the CSM at all scales. No time-variable optical absorption lines are apparent in SN\,2011fe \citep{Patat_etal11b}, consistent with an absence of CSM. We note that preliminary statistics hint that roughly half of SNe Ia occur in similarly ``clean" environments, while $\sim$25\% of SNe Ia in spiral galaxies show narrow blue-shifted absorption lines \citep{Sternberg_etal11}. If all SN Ia systems that show blue-shifted absorption lines have recurrent nova progenitors, this would conflict with our census of recurrent novae in the local universe, predicting many more nova explosions than are observed \citep{dellaValle_Livio96}.

We emphasize, however, that the progenitors of SNe Ia may be diverse, and constraints placed on the progenitor of SN\,2011fe will not necessarily hold for the entire class of SNe Ia. For example, while most observations of SN Ia remnants are not consistent with accretion winds \citep{Badenes_etal07}, this does not imply that all SN Ia progenitor systems do not host accretion winds. For one, RCW 86 may provide a counter example, as recent work shows that it is likely expanding into a wind-blown cavity \citep{Williams_etal11}; however, it remains uncertain if RCW 86 is a SN Ia or core-collapse remnant. Also, many SNe Ia do not show time-variable Na I D absorption lines \citep{Patat_etal07b, Simon_etal07, Sternberg_etal11} including SN\,2011fe \citep{Patat_etal11b}.  This overall diversity could be due to viewing angle effects, differences in metallicity of the CSM, or a range of circumbinary environments in the progenitor system.

\subsection{Directly Detecting the Binary Companions of SNe Ia}
While it is possible to directly detect the massive progenitor stars of core-collapse SNe in nearby galaxies using archival pre-explosion optical images \citep{Smartt09}, both the massive WDs that produce SNe Ia and their binary companions are too faint to be detected in external galaxies \citep{Maoz_Mannucci08}.  Indeed, no optical source was detected pre-explosion at the position of SN\,2011fe \citep{Li_etal11b}. However, in our Galaxy, we can search Type Ia SNRs for stars with unusual proper motions and abundances, as might be expected for the binary companion of the now-destroyed WD. One such star was proposed to be the sub-giant star G in the Tycho SNR \citep{Ruiz-Lapuente_etal04}, but more recently star G's affililation with the SNR has been called into question \citep{Kerzendorf_etal09}. Additional studies are required, of Tycho and other SNRs in the Milky Way and Magellanic Clouds, to directly search for binary companions to SNe Ia.

When the SN shockwave plows over a non-degenerate companion, this interaction may leave an observable mark upon the SN Ia explosion itself. If the binary has a relatively small separation, some of the companion star's envelope should be entrained in the SN ejecta, and become detectable in late-time nebular spectra as H$\alpha$ emission \citep{Marietta_etal00}. Deep limits have been placed on the presence of such emission, implying that $< 0.01$ M$_{\odot}$ of hydrogen-rich material is swept from a companion star, and presenting a challenge to progenitor scenarios where the companion is filling its Roche lobe \citep{Leonard07}.  The interaction may also be detectable as an early-time blue component in the optical SN light curve \citep{Kasen10}; the magnitude of this early component  will depend on viewing angle and the nature of the companion, with Roche-lobe filling red giant stars producing the brightest signature. In the significant samples of SNe Ia now available, this effect has never been detected, implying that $<$20\% of SNe Ia have red giant companions \citep{Hayden_etal10,Bianco_etal11,Brown_etal12}. In addition, no blue early-time blue component was detected for SN\,2011fe, placing strong constraints on possible Roche-lobe filling companions ($<$1 M$_{\odot}$ red giant, or $<$3 M$_{\odot}$ main sequence; \citealt{Brown_etal11}; see also \citealt{Nugent_etal11,Bloom_etal12}).

\section{Conclusions}

Using deep EVLA radio observations, we have shown that the progenitor system of SN\,2011fe did not host a red giant secondary or a wind-producing high accretion rate. We also rule out a significant fraction of the parameter space for stably burning WD or recurrent nova progenitors---constraints that are bolstered by pre-explosion \emph{Chandra} and {HST} imaging and a lack of an early-time blue component in the optical light curve.

Our study may be viewed as consistent with the growing body of evidence favoring DD progenitors, which includes measurements of the delay time distribution of SNe Ia \citep{Maoz_etal10} and computational successes with theoretical DD explosions \citep{Pakmor_etal10}.  However, a recent preliminary model of a WD--WD merger suggests the presence of dense circumbinary medium \cite{Shen_etal11}, which could be constrained by these and future radio observations.  A thorough treatment of this scenario would require a re-evaluation of the SN shockwave dynamics, as it is predicted that a dense envelope of material will surround the primary WD at small radius ($\sim$0.1 M$_{\odot}$ at $\lesssim 10^{13}$ cm), which would slow the shockwave. At larger radius, the WD is surrounded by a $\rho_{\rm CSM} \propto r^{-2}$ wind profile, expelling roughly $\sim$0.1 M$_{\odot}$ over $10^{4}$ years at a velocity of $\sim$100 km s$^{-1}$, producing a relatively high $\dot{M}/v_{w} \approx 10^{-5} {{{\rm M}_{\odot}\ {\rm yr}^{-1}} \over {{\rm 100\ km\ s}^{-1}}}$. It is the interaction between this wind and the SN shockwave that would likely produce detectable radio emission, but more detailed models of the density profile and mass loss in DD systems are needed to understand the velocity evolution of the shockwave and the predicted synchrotron signal. 

More exotic scenarios, like spin-up/spin-down models \citep{Justham11,diStefano_etal11} or the core-degenerate scenario \citep{Ilkov_Soker12}, with a significant delay between mass loss and explosion ($\gtrsim 10^5$ years) are also not excluded. While our EVLA upper limits place the most stringent constraints to date on the CSM density around a SN Ia, and our analysis indicates that most SD progenitor models are ruled out, detailed studies (theoretical and observational) of the circumbinary environments surrounding accreting WDs are required to further shed light on their connection to Type Ia SNe.

\acknowledgements
We are grateful to Sumner Starrfield and an anonymous referee for their insights. The EVLA is operated by the National Radio Astronomy Observatory, a facility of the National Science Foundation operated under cooperative agreement by Associated Universities, Inc. L.C. is a Jansky Fellow of the National Radio Astronomy Observatory. This work made use of THINGS, `The HI Nearby Galaxy Survey' (Walter et al. 2008). This research has also made use of the NASA/IPAC Extragalactic Database (NED) which is operated by the Jet Propulsion Laboratory, California Institute of Technology, under contract with the National Aeronautics and Space Administration. RAC acknowledges support from the NSF under grant AST-0807727.

\bibliographystyle{apj}

\bibliography{refs}

\begin{thebibliography}{99}
\expandafter\ifx\csname natexlab\endcsname\relax\def\natexlab#1{#1}\fi

\bibitem[{{Badenes} {et~al.}(2006){Badenes}, {Borkowski}, {Hughes}, {Hwang}, \&
  {Bravo}}]{Badenes_etal06}
{Badenes}, C., {Borkowski}, K.~J., {Hughes}, J.~P., {Hwang}, U., \& {Bravo}, E.
  2006, \apj, 645, 1373

\bibitem[{{Badenes} {et~al.}(2007){Badenes}, {Hughes}, {Bravo}, \&
  {Langer}}]{Badenes_etal07}
{Badenes}, C., {Hughes}, J.~P., {Bravo}, E., \& {Langer}, N. 2007, \apj, 662,
  472

\bibitem[{{Badenes} {et~al.}(2008){Badenes}, {Hughes}, {Cassam-Chena{\"i}}, \&
  {Bravo}}]{Badenes_etal08}
{Badenes}, C., {Hughes}, J.~P., {Cassam-Chena{\"i}}, G., \& {Bravo}, E. 2008,
  \apj, 680, 1149

\bibitem[{{Berger} {et~al.}(2002){Berger}, {Kulkarni}, \&
  {Chevalier}}]{Berger_etal02}
{Berger}, E., {Kulkarni}, S.~R., \& {Chevalier}, R.~A. 2002, \apj, 577, L5

\bibitem[{{Berger} {et~al.}(2003){Berger}, {Kulkarni}, {Frail}, \&
  {Soderberg}}]{Berger_etal03}
{Berger}, E., {Kulkarni}, S.~R., {Frail}, D.~A., \& {Soderberg}, A.~M. 2003,
  \apj, 599, 408

\bibitem[{{Bianco} {et~al.}(2011){Bianco}, {Howell}, {Sullivan}, {Conley},
  {Kasen}, {Gonz{\'a}lez-Gait{\'a}n}, {Guy}, {Astier}, {Balland}, {Carlberg},
  {Fouchez}, {Fourmanoit}, {Hardin}, {Hook}, {Lidman}, {Pain},
  {Palanque-Delabrouille}, {Perlmutter}, {Perrett}, {Pritchet}, {Regnault},
  {Rich}, \& {Ruhlmann-Kleider}}]{Bianco_etal11}
{Bianco}, F.~B., {Howell}, D.~A., {Sullivan}, M., {et~al.} 2011, \apj, 741, 20

\bibitem[{{Blondin} {et~al.}(2009){Blondin}, {Prieto}, {Patat}, {Challis},
  {Hicken}, {Kirshner}, {Matheson}, \& {Modjaz}}]{Blondin_etal09}
{Blondin}, S., {Prieto}, J.~L., {Patat}, F., {et~al.} 2009, \apj, 693, 207

\bibitem[{{Bloom} {et~al.}(2012){Bloom}, {Kasen}, {Shen}, {Nugent}, {Butler},
  {Graham}, {Howell}, {Kolb}, {Holmes}, {Haswell}, {Burwitz}, {Rodriguez}, \&
  {Sullivan}}]{Bloom_etal12}
{Bloom}, J.~S., {Kasen}, D., {Shen}, K.~J., {et~al.} 2012, \apjl, 744, L17

\bibitem[{{Borkowski} {et~al.}(2010){Borkowski}, {Reynolds}, {Green}, {Hwang},
  {Petre}, {Krishnamurthy}, \& {Willett}}]{Borkowski_etal10}
{Borkowski}, K.~J., {Reynolds}, S.~P., {Green}, D.~A., {et~al.} 2010, \apjl,
  724, L161

\bibitem[{{Brown} {et~al.}(2012){Brown}, {Dawson}, {Harris}, {Olmstead},
  {Milne}, \& {Roming}}]{Brown_etal12}
{Brown}, P.~J., {Dawson}, K.~S., {Harris}, D.~W., {et~al.} 2012, \apj, in press

\bibitem[{{Brown} {et~al.}(2011){Brown}, {Dawson}, {de Pasquale}, {Gronwall},
  {Holland}, {Immler}, {Kuin}, {Mazzali}, {Milne}, {Oates}, \&
  {Siegel}}]{Brown_etal11}
{Brown}, P.~J., {Dawson}, K.~S., {de Pasquale}, M., {et~al.} 2011, arXiv
  1110.2538

\bibitem[{{Carlton} {et~al.}(2011){Carlton}, {Borkowski}, {Reynolds}, {Hwang},
  {Petre}, {Green}, {Krishnamurthy}, \& {Willett}}]{Carlton_etal11}
{Carlton}, A.~K., {Borkowski}, K.~J., {Reynolds}, S.~P., {et~al.} 2011, \apjl,
  737, L22

\bibitem[{{Chen} {et~al.}(2011){Chen}, {Han}, \& {Tout}}]{Chen_etal11}
{Chen}, X., {Han}, Z., \& {Tout}, C.~A. 2011, \apj, 735, L31+

\bibitem[{{Chevalier}(1982{\natexlab{a}})}]{Chevalier82}
{Chevalier}, R.~A. 1982{\natexlab{a}}, \apj, 258, 790

\bibitem[{{Chevalier}(1982{\natexlab{b}})}]{Chevalier82b}
---. 1982{\natexlab{b}}, \apj, 259, 302

\bibitem[{{Chevalier}(1998)}]{c98}
---. 1998, \apj, 499, 810

\bibitem[{{Chevalier} \& {Fransson}(2006)}]{cf06}
{Chevalier}, R.~A., \& {Fransson}, C. 2006, \apj, 651, 381

\bibitem[{{Chomiuk} \& {Soderberg}(2011)}]{Chomiuk_Soderberg11}
{Chomiuk}, L., \& {Soderberg}, A. 2011, ATel, 3597, 1

\bibitem[{{Chugai}(2008)}]{Chugai08}
{Chugai}, N.~N. 2008, Astronomy Letters, 34, 389

\bibitem[{{Colgate}(1970)}]{Colgate70}
{Colgate}, S.~A. 1970, in Proc.~of the 11th International Cosmic Ray
  Conference, 353

\bibitem[{{Cowley} {et~al.}(1998){Cowley}, {Schmidtke}, {Crampton}, \&
  {Hutchings}}]{Cowley_etal98}
{Cowley}, A.~P., {Schmidtke}, P.~C., {Crampton}, D., \& {Hutchings}, J.~B.
  1998, \apj, 504, 854

\bibitem[{{Dale} {et~al.}(2009){Dale}, {Cohen}, {Johnson}, {Schuster},
  {Calzetti}, {Engelbracht}, {Gil de Paz}, {Kennicutt}, {Lee}, {Begum},
  {Block}, {Dalcanton}, {Funes}, {Gordon}, {Johnson}, {Marble}, {Sakai},
  {Skillman}, {van Zee}, {Walter}, {Weisz}, {Williams}, {Wu}, \&
  {Wu}}]{Dale_etal09}
{Dale}, D.~A., {Cohen}, S.~A., {Johnson}, L.~C., {et~al.} 2009, \apj, 703, 517

\bibitem[{{della Valle} \& {Livio}(1996)}]{dellaValle_Livio96}
{della Valle}, M., \& {Livio}, M. 1996, \apj, 473, 240

\bibitem[{{Deufel} {et~al.}(1999){Deufel}, {Barwig}, {{\v S}imi{\'c} }, {Wolf},
  \& {Drory}}]{Deufel_etal99}
{Deufel}, B., {Barwig}, H., {{\v S}imi{\'c} }, D., {Wolf}, S., \& {Drory}, N.
  1999, \aap, 343, 455

\bibitem[{{Di Stefano} {et~al.}(2011){Di Stefano}, {Voss}, \&
  {Claeys}}]{diStefano_etal11}
{Di Stefano}, R., {Voss}, R., \& {Claeys}, J.~S.~W. 2011, \apj, 738, L1

\bibitem[{{Diaz} {et~al.}(2010){Diaz}, {Williams}, {Luna}, {Moraes}, \&
  {Takeda}}]{Diaz_etal10}
{Diaz}, M.~P., {Williams}, R.~E., {Luna}, G.~J., {Moraes}, M., \& {Takeda}, L.
  2010, \aj, 140, 1860

\bibitem[{{Duschl} {et~al.}(1990){Duschl}, {Livio}, \&
  {Truran}}]{Duschl_etal90}
{Duschl}, W.~J., {Livio}, M., \& {Truran}, J.~W. 1990, \apj, 360, 232

\bibitem[{{Evans} {et~al.}(2001){Evans}, {Krautter}, {Vanzi}, \&
  {Starrfield}}]{Evans_etal01}
{Evans}, A., {Krautter}, J., {Vanzi}, L., \& {Starrfield}, S. 2001, \aap, 378,
  132

\bibitem[{{Ferri{\`e}re}(2001)}]{Ferriere01}
{Ferri{\`e}re}, K.~M. 2001, Reviews of Modern Physics, 73, 1031

\bibitem[{{Fruchter} {et~al.}(2006){Fruchter}, {Levan}, {Strolger},
  {Vreeswijk}, {Thorsett}, {Bersier}, {Burud}, {Castro Cer{\'o}n},
  {Castro-Tirado}, {Conselice}, {Dahlen}, {Ferguson}, {Fynbo}, {Garnavich},
  {Gibbons}, {Gorosabel}, {Gull}, {Hjorth}, {Holland}, {Kouveliotou}, {Levay},
  {Livio}, {Metzger}, {Nugent}, {Petro}, {Pian}, {Rhoads}, {Riess}, {Sahu},
  {Smette}, {Tanvir}, {Wijers}, \& {Woosley}}]{Fruchter_etal06}
{Fruchter}, A.~S., {Levan}, A.~J., {Strolger}, L., {et~al.} 2006, Nature, 441,
  463

\bibitem[{{Hachisu} {et~al.}(2000){Hachisu}, {Kato}, {Kato}, \&
  {Matsumoto}}]{Hachisu_etal00}
{Hachisu}, I., {Kato}, M., {Kato}, T., \& {Matsumoto}, K. 2000, \apj, 528, L97

\bibitem[{{Hachisu} {et~al.}(1999){Hachisu}, {Kato}, \&
  {Nomoto}}]{Hachisu_etal99}
{Hachisu}, I., {Kato}, M., \& {Nomoto}, K. 1999, \apj, 522, 487

\bibitem[{{Han} \& {Podsiadlowski}(2004)}]{Han_Podsiadlowski04}
{Han}, Z., \& {Podsiadlowski}, P. 2004, \mnras, 350, 1301

\bibitem[{{Hancock} {et~al.}(2011){Hancock}, {Gaensler}, \&
  {Murphy}}]{Hancock_etal11}
{Hancock}, P.~P., {Gaensler}, B.~M., \& {Murphy}, T. 2011, \apjl, 735, L35

\bibitem[{{Hayden} {et~al.}(2010){Hayden}, {Garnavich}, {Kasen}, {Dilday},
  {Frieman}, {Jha}, {Lampeitl}, {Nichol}, {Sako}, {Schneider}, {Smith},
  {Sollerman}, \& {Wheeler}}]{Hayden_etal10}
{Hayden}, B.~T., {Garnavich}, P.~M., {Kasen}, D., {et~al.} 2010, \apj, 722,
  1691

\bibitem[{{Hillebrandt} \& {Niemeyer}(2000)}]{Hillebrandt_Niemeyer00}
{Hillebrandt}, W., \& {Niemeyer}, J.~C. 2000, \araa, 38, 191

\bibitem[{{Hoopes} {et~al.}(2001){Hoopes}, {Walterbos}, \& {Bothun}}]{hwb01}
{Hoopes}, C.~G., {Walterbos}, R.~A.~M., \& {Bothun}, G.~D. 2001, \apj, 559, 878

\bibitem[{{Horesh} {et~al.}(2012){Horesh}, {Kulkarni}, {Fox}, {Carpenter},
  {Kasliwal}, {Ofek}, {Quimby}, {Gal-Yam}, {Cenko}, {de Bruyn}, {Kamble},
  {Wijers}, {van der Horst}, {Kouveliotou}, {Podsiadlowski}, {Sullivan},
  {Maguire}, {Howell}, {Nugent}, {Gehrels}, {Law}, {Poznanski}, \&
  {Shara}}]{Horesh_etal11}
{Horesh}, A., {Kulkarni}, S.~R., {Fox}, D.~B., {et~al.} 2012, \apj, 746, 21

\bibitem[{{Huang} \& {Yu}(1996)}]{Huang_Yu96}
{Huang}, R.-Q., \& {Yu}, K.~N. 1996, Chinese Journal of Astronomy \&
  Astrophysics, 20, 175

\bibitem[{{Hughes} {et~al.}(2007){Hughes}, {Chugai}, {Chevalier}, {Lundqvist},
  \& {Schlegel}}]{Hughes_etal07}
{Hughes}, J.~P., {Chugai}, N., {Chevalier}, R., {Lundqvist}, P., \& {Schlegel},
  E. 2007, \apj, 670, 1260

\bibitem[{{Iben} \& {Tutukov}(1984)}]{Iben_Tutukov84}
{Iben}, Jr., I., \& {Tutukov}, A.~V. 1984, \apjs, 54, 335

\bibitem[{{Ilkov} \& {Soker}(2012)}]{Ilkov_Soker12}
{Ilkov}, M., \& {Soker}, N. 2012, \mnras, 419, 1695

\bibitem[{{Immler} {et~al.}(2006){Immler}, {Brown}, {Milne}, {The}, {Petre},
  {Gehrels}, {Burrows}, {Nousek}, {Williams}, {Pian}, {Mazzali}, {Nomoto},
  {Chevalier}, {Mangano}, {Holland}, {Roming}, {Greiner}, \&
  {Pooley}}]{Immler_etal06}
{Immler}, S., {Brown}, P.~J., {Milne}, P., {et~al.} 2006, \apjl, 648, L119

\bibitem[{{Justham}(2011)}]{Justham11}
{Justham}, S. 2011, \apj, 730, L34+

\bibitem[{{Kafka} \& {Williams}(2011)}]{Kafka_Williams11}
{Kafka}, S., \& {Williams}, R. 2011, \aap, 526, A83

\bibitem[{{Kasen}(2010)}]{Kasen10}
{Kasen}, D. 2010, \apj, 708, 1025

\bibitem[{{Kelly} {et~al.}(2008){Kelly}, {Kirshner}, \& {Pahre}}]{Kelly_etal08}
{Kelly}, P.~L., {Kirshner}, R.~P., \& {Pahre}, M. 2008, \apj, 687, 1201

\bibitem[{{Kerzendorf} {et~al.}(2009){Kerzendorf}, {Schmidt}, {Asplund},
  {Nomoto}, {Podsiadlowski}, {Frebel}, {Fesen}, \& {Yong}}]{Kerzendorf_etal09}
{Kerzendorf}, W.~E., {Schmidt}, B.~P., {Asplund}, M., {et~al.} 2009, \apj, 701,
  1665

\bibitem[{{Law} {et~al.}(2009){Law}, {Kulkarni}, {Dekany}, {Ofek}, {Quimby},
  {Nugent}, {Surace}, {Grillmair}, {Bloom}, {Kasliwal}, {Bildsten}, {Brown},
  {Cenko}, {Ciardi}, {Croner}, {Djorgovski}, {van Eyken}, {Filippenko}, {Fox},
  {Gal-Yam}, {Hale}, {Hamam}, {Helou}, {Henning}, {Howell}, {Jacobsen},
  {Laher}, {Mattingly}, {McKenna}, {Pickles}, {Poznanski}, {Rahmer}, {Rau},
  {Rosing}, {Shara}, {Smith}, {Starr}, {Sullivan}, {Velur}, {Walters}, \&
  {Zolkower}}]{Law_etal09}
{Law}, N.~M., {Kulkarni}, S.~R., {Dekany}, R.~G., {et~al.} 2009, \pasp, 121,
  1395

\bibitem[{{Leonard}(2007)}]{Leonard07}
{Leonard}, D.~C. 2007, \apj, 670, 1275

\bibitem[{{Li} {et~al.}(2011{\natexlab{a}}){Li}, {Chornock}, {Leaman},
  {Filippenko}, {Poznanski}, {Wang}, {Ganeshalingam}, \&
  {Mannucci}}]{Li_etal11a}
{Li}, W., {Chornock}, R., {Leaman}, J., {et~al.} 2011{\natexlab{a}}, \mnras,
  412, 1473

\bibitem[{{Li} {et~al.}(2011{\natexlab{b}}){Li}, {Bloom}, {Podsiadlowski},
  {Miller}, {Cenko}, {Jha}, {Sullivan}, {Howell}, {Nugent}, {Butler}, {Ofek},
  {Kasliwal}, {Richards}, {Stockton}, {Shih}, {Bildsten}, {Shara}, {Bibby},
  {Filippenko}, {Ganeshalingam}, {Silverman}, {Kulkarni}, {Law}, {Poznanski},
  {Quimby}, {McCully}, {Patel}, {Maguire}, \& {Shen}}]{Li_etal11b}
{Li}, W., {Bloom}, J.~S., {Podsiadlowski}, P., {et~al.} 2011{\natexlab{b}},
  \nat, 480, 348

\bibitem[{{Liu} {et~al.}(2011){Liu}, {Di Stefano}, {Wang}, \&
  {Moe}}]{Liu_etal11}
{Liu}, J., {Di Stefano}, R., {Wang}, T., \& {Moe}, M. 2011, arXiv 1110.2506

\bibitem[{{Maoz} \& {Mannucci}(2008)}]{Maoz_Mannucci08}
{Maoz}, D., \& {Mannucci}, F. 2008, \mnras, 388, 421

\bibitem[{{Maoz} {et~al.}(2010){Maoz}, {Sharon}, \& {Gal-Yam}}]{Maoz_etal10}
{Maoz}, D., {Sharon}, K., \& {Gal-Yam}, A. 2010, \apj, 722, 1879

\bibitem[{{Margutti} {et~al.}(2012){Margutti}, {Soderberg}, {Chomiuk},
  {Chevalier}, {Hurley}, {Milisavljevic}, {Foley}, {Hughes}, {Slane},
  {Fransson}, {Moe}, {Barthelmy}, {Boynton}, {Briggs}, {Connaughton}, {Costa},
  {Cummings}, {Del Monte}, {Enos}, {Fellows}, {Feroci}, {Fukazawa}, {Gehrels},
  {Goldsten}, {Golovin}, {Hanabata}, {Harshman}, {Krimm}, {Litvak},
  {Makishima}, {Marisaldi}, {Mitrofanov}, {Murakami}, {Ohno}, {Palmer},
  {Sanin}, {Starr}, \& {Svinkin}}]{Margutti_etal11}
{Margutti}, R., {Soderberg}, A.~M., {Chomiuk}, L., {et~al.} 2012, arXiv
  1202.0741

\bibitem[{{Marietta} {et~al.}(2000){Marietta}, {Burrows}, \&
  {Fryxell}}]{Marietta_etal00}
{Marietta}, E., {Burrows}, A., \& {Fryxell}, B. 2000, \apjs, 128, 615

\bibitem[{{Mattila} {et~al.}(2005){Mattila}, {Lundqvist}, {Sollerman}, {Kozma},
  {Baron}, {Fransson}, {Leibundgut}, \& {Nomoto}}]{Mattila_etal05}
{Mattila}, S., {Lundqvist}, P., {Sollerman}, J., {et~al.} 2005, \aap, 443, 649

\bibitem[{{Matzner} \& {McKee}(1999)}]{mm99}
{Matzner}, C.~D., \& {McKee}, C.~F. 1999, \apj, 510, 379

\bibitem[{{Mazzali} {et~al.}(2007){Mazzali}, {R{\"o}pke}, {Benetti}, \&
  {Hillebrandt}}]{Mazzali_etal07}
{Mazzali}, P.~A., {R{\"o}pke}, F.~K., {Benetti}, S., \& {Hillebrandt}, W. 2007,
  Science, 315, 825

\bibitem[{{Mohamed} \& {Podsiadlowski}(2011)}]{Mohamed_Podsiadlowski11}
{Mohamed}, S., \& {Podsiadlowski}, P. 2011, in Asymmetric Planetary Nebulae 5
  Conf., ed. {A.~Zijlstra, F.~Lykou, I.~McDonald, \& E.~Lagadec}

\bibitem[{{Murray}(2002)}]{Murray02}
{Murray}, N. 2002, in ASP Conf.~Ser., Vol. 261, The Physics of Cataclysmic
  Variables and Related Objects, ed. {B.~T.~G{\"a}nsicke, K.~Beuermann, \&
  K.~Reinsch}, 308

\bibitem[{{Nakar} \& {Sari}(2012)}]{Nakar_Sari11}
{Nakar}, E., \& {Sari}, R. 2012, \apj, 747, 88

\bibitem[{{Nomoto}(1982)}]{Nomoto82}
{Nomoto}, K. 1982, \apj, 253, 798

\bibitem[{{Nomoto} {et~al.}(2007){Nomoto}, {Saio}, {Kato}, \&
  {Hachisu}}]{Nomoto_etal07}
{Nomoto}, K., {Saio}, H., {Kato}, M., \& {Hachisu}, I. 2007, \apj, 663, 1269

\bibitem[{{Nugent} {et~al.}(2011){Nugent}, {Sullivan}, {Cenko}, {Thomas},
  {Kasen}, {Howell}, {Bersier}, {Bloom}, {Kulkarni}, {Kandrashoff},
  {Filippenko}, {Silverman}, {Marcy}, {Howard}, {Isaacson}, {Maguire},
  {Suzuki}, {Tarlton}, {Pan}, {Bildsten}, {Fulton}, {Parrent}, {Sand},
  {Podsiadlowski}, {Bianco}, {Dilday}, {Graham}, {Lyman}, {James}, {Kasliwal},
  {Law}, {Quimby}, {Hook}, {Walker}, {Mazzali}, {Pian}, {Ofek}, {Gal-Yam}, \&
  {Poznanski}}]{Nugent_etal11}
{Nugent}, P.~E., {Sullivan}, M., {Cenko}, S.~B., {et~al.} 2011, \nat, 480, 344

\bibitem[{{Pakmor} {et~al.}(2010){Pakmor}, {Kromer}, {R{\"o}pke}, {Sim},
  {Ruiter}, \& {Hillebrandt}}]{Pakmor_etal10}
{Pakmor}, R., {Kromer}, M., {R{\"o}pke}, F.~K., {et~al.} 2010, Nature, 463, 61

\bibitem[{{Panagia} {et~al.}(2006){Panagia}, {Van Dyk}, {Weiler}, {Sramek},
  {Stockdale}, \& {Murata}}]{Panagia_etal06}
{Panagia}, N., {Van Dyk}, S.~D., {Weiler}, K.~W., {et~al.} 2006, \apj, 646, 369

\bibitem[{{Patat} {et~al.}(2011{\natexlab{a}}){Patat}, {Chugai},
  {Podsiadlowski}, {Mason}, {Melo}, \& {Pasquini}}]{Patat_etal11a}
{Patat}, F., {Chugai}, N.~N., {Podsiadlowski}, P., {et~al.} 2011{\natexlab{a}},
  \aap, 530, A63+

\bibitem[{{Patat} {et~al.}(2007{\natexlab{a}}){Patat}, {Chandra}, {Chevalier},
  {Justham}, {Podsiadlowski}, {Wolf}, {Gal-Yam}, {Pasquini}, {Crawford},
  {Mazzali}, {Pauldrach}, {Nomoto}, {Benetti}, {Cappellaro}, {Elias-Rosa},
  {Hillebrandt}, {Leonard}, {Pastorello}, {Renzini}, {Sabbadin}, {Simon}, \&
  {Turatto}}]{Patat_etal07}
{Patat}, F., {Chandra}, P., {Chevalier}, R., {et~al.} 2007{\natexlab{a}},
  Science, 317, 924

\bibitem[{{Patat} {et~al.}(2007{\natexlab{b}}){Patat}, {Benetti}, {Justham},
  {Mazzali}, {Pasquini}, {Cappellaro}, {Della Valle}, {-Podsiadlowski},
  {Turatto}, {Gal-Yam}, \& {Simon}}]{Patat_etal07b}
{Patat}, F., {Benetti}, S., {Justham}, S., {et~al.} 2007{\natexlab{b}}, \aap,
  474, 931

\bibitem[{{Patat} {et~al.}(2011{\natexlab{b}}){Patat}, {Cordiner}, {Cox},
  {Anderson}, {Harutyunyan}, {Kotak}, {Palaversa}, {Stanishev}, {Tomasella},
  {Benetti}, {Goobar}, {Pastorello}, \& {Sollerman}}]{Patat_etal11b}
{Patat}, F., {Cordiner}, M.~A., {Cox}, N.~L.~J., {et~al.} 2011{\natexlab{b}},
  arXiv 1112.0247

\bibitem[{{Perley} {et~al.}(2011){Perley}, {Chandler}, {Butler}, \&
  {Wrobel}}]{Perley_etal11}
{Perley}, R.~A., {Chandler}, C.~J., {Butler}, B.~J., \& {Wrobel}, J.~M. 2011,
  \apj, 739, L1+

\bibitem[{{Raymond} {et~al.}(2007){Raymond}, {Korreck}, {Sedlacek}, {Blair},
  {Ghavamian}, \& {Sankrit}}]{Raymond_etal07}
{Raymond}, J.~C., {Korreck}, K.~E., {Sedlacek}, Q.~C., {et~al.} 2007, \apj,
  659, 1257

\bibitem[{{Readhead}(1994)}]{Readhead94}
{Readhead}, A.~C.~S. 1994, \apj, 426, 51

\bibitem[{{Reynolds} {et~al.}(2008){Reynolds}, {Borkowski}, {Green}, {Hwang},
  {Harrus}, \& {Petre}}]{Reynolds_etal08}
{Reynolds}, S.~P., {Borkowski}, K.~J., {Green}, D.~A., {et~al.} 2008, \apjl,
  680, L41

\bibitem[{{Ruiz-Lapuente} {et~al.}(2004){Ruiz-Lapuente}, {Comeron},
  {M{\'e}ndez}, {Canal}, {Smartt}, {Filippenko}, {Kurucz}, {Chornock}, {Foley},
  {Stanishev}, \& {Ibata}}]{Ruiz-Lapuente_etal04}
{Ruiz-Lapuente}, P., {Comeron}, F., {M{\'e}ndez}, J., {et~al.} 2004, Nature,
  431, 1069

\bibitem[{{Russell} \& {Immler}(2012)}]{Russell_Immler12}
{Russell}, B.~R., \& {Immler}, S. 2012, \apj, in press

\bibitem[{{Schaefer}(2010)}]{Schaefer10}
{Schaefer}, B.~E. 2010, \apjs, 187, 275

\bibitem[{{Seaquist} \& {Taylor}(1990)}]{Seaquist_Taylor90}
{Seaquist}, E.~R., \& {Taylor}, A.~R. 1990, \apj, 349, 313

\bibitem[{{Shappee} \& {Stanek}(2011)}]{ss11}
{Shappee}, B.~J., \& {Stanek}, K.~Z. 2011, \apj, 733, 124

\bibitem[{{Shen} \& {Bildsten}(2007)}]{Shen_Bildsten07}
{Shen}, K.~J., \& {Bildsten}, L. 2007, \apj, 660, 1444

\bibitem[{{Shen} \& {Bildsten}(2009)}]{Shen_Bildsten09}
---. 2009, \apj, 692, 324

\bibitem[{{Shen} {et~al.}(2011){Shen}, {Bildsten}, {Kasen}, \&
  {Quataert}}]{Shen_etal11}
{Shen}, K.~J., {Bildsten}, L., {Kasen}, D., \& {Quataert}, E. 2011, arXiv
  1108.4036

\bibitem[{{Simon} {et~al.}(2007){Simon}, {Gal-Yam}, {Penprase}, {Li}, {Quimby},
  {Silverman}, {Allende Prieto}, {Wheeler}, {Filippenko}, {Martinez}, {Beeler},
  \& {Patat}}]{Simon_etal07}
{Simon}, J.~D., {Gal-Yam}, A., {Penprase}, B.~E., {et~al.} 2007, \apj, 671, L25

\bibitem[{{Simon} {et~al.}(2009){Simon}, {Gal-Yam}, {Gnat}, {Quimby},
  {Ganeshalingam}, {Silverman}, {Blondin}, {Li}, {Filippenko}, {Wheeler},
  {Kirshner}, {Patat}, {Nugent}, {Foley}, {Vogt}, {Butler}, {Peek},
  {Rosolowsky}, {Herczeg}, {Sauer}, \& {Mazzali}}]{Simon_etal09}
{Simon}, J.~D., {Gal-Yam}, A., {Gnat}, O., {et~al.} 2009, \apj, 702, 1157

\bibitem[{{Smartt}(2009)}]{Smartt09}
{Smartt}, S.~J. 2009, \araa, 47, 63

\bibitem[{{Soderberg} {et~al.}(2010){Soderberg}, {Chakraborti}, {Pignata},
  {Chevalier}, {Chandra}, {Ray}, {Wieringa}, {Copete}, {Chaplin},
  {Connaughton}, {Barthelmy}, {Bietenholz}, {Chugai}, {Stritzinger}, {Hamuy},
  {Fransson}, {Fox}, {Levesque}, {Grindlay}, {Challis}, {Foley}, {Kirshner},
  {Milne}, \& {Torres}}]{Soderberg_etal10}
{Soderberg}, A.~M., {Chakraborti}, S., {Pignata}, G., {et~al.} 2010, \nat, 463,
  513

\bibitem[{{Soderberg} {et~al.}(2011){Soderberg}, {Margutti}, {Zauderer},
  {Krauss}, {Katz}, {Chomiuk}, {Dittmann}, {Nakar}, {Sakamoto}, {Kawai},
  {Hurley}, {Barthelmy}, {Toizumi}, {Morii}, {Chevalier}, {Gurwell},
  {Petitpas}, {Rupen}, {Alexander}, {Levesque}, {Fransson}, {Brunthaler},
  {Bietenholz}, {Chugai}, {Connaughton}, {Briggs}, {Meegan}, {von Kienlin},
  {Zhang}, {Rau}, {Golenetskii}, {Mazets}, \& {Cline}}]{Soderberg_etal11}
{Soderberg}, A.~M., {Margutti}, R., {Zauderer}, B.~A., {et~al.} 2011, arXiv
  1107.1876

\bibitem[{{Sternberg} {et~al.}(2011){Sternberg}, {Gal-Yam}, {Simon}, {Leonard},
  {Quimby}, {Phillips}, {Morrell}, {Thompson}, {Ivans}, {Marshall},
  {Filippenko}, {Marcy}, {Bloom}, {Patat}, {Foley}, {Yong}, {Penprase},
  {Beeler}, {Allende Prieto}, \& {Stringfellow}}]{Sternberg_etal11}
{Sternberg}, A., {Gal-Yam}, A., {Simon}, J.~D., {et~al.} 2011, Science, 333,
  856

\bibitem[{{Walter} {et~al.}(2008){Walter}, {Brinks}, {de Blok}, {Bigiel},
  {Kennicutt}, {Thornley}, \& {Leroy}}]{Walter_etal08}
{Walter}, F., {Brinks}, E., {de Blok}, W.~J.~G., {et~al.} 2008, \aj, 136, 2563

\bibitem[{{Webbink}(1984)}]{Webbink84}
{Webbink}, R.~F. 1984, \apj, 277, 355

\bibitem[{{Weiler} {et~al.}(2002){Weiler}, {Panagia}, {Montes}, \&
  {Sramek}}]{Weiler_etal02}
{Weiler}, K.~W., {Panagia}, N., {Montes}, M.~J., \& {Sramek}, R.~A. 2002,
  \araa, 40, 387

\bibitem[{{Whelan} \& {Iben}(1973)}]{Whelan_Iben73}
{Whelan}, J., \& {Iben}, Jr., I. 1973, \apj, 186, 1007

\bibitem[{{Williams} {et~al.}(2011){Williams}, {Blair}, {Blondin}, {Borkowski},
  {Ghavamian}, {Long}, {Raymond}, {Reynolds}, {Rho}, \&
  {Winkler}}]{Williams_etal11}
{Williams}, B.~J., {Blair}, W.~P., {Blondin}, J.~M., {et~al.} 2011, \apj, 741,
  96

\bibitem[{{Williams} {et~al.}(2008){Williams}, {Mason}, {Della Valle}, \&
  {Ederoclite}}]{Williams_etal08}
{Williams}, R., {Mason}, E., {Della Valle}, M., \& {Ederoclite}, A. 2008, \apj,
  685, 451

\bibitem[{{Wood-Vasey} \& {Sokoloski}(2006)}]{Wood-Vasey_Sokoloski06}
{Wood-Vasey}, W.~M., \& {Sokoloski}, J.~L. 2006, \apj, 645, L53

\bibitem[{{Yaron} {et~al.}(2005){Yaron}, {Prialnik}, {Shara}, \&
  {Kovetz}}]{Yaron_etal05}
{Yaron}, O., {Prialnik}, D., {Shara}, M.~M., \& {Kovetz}, A. 2005, \apj, 623,
  398

\bibitem[{{Zhang} {et~al.}(2009){Zhang}, {MacFadyen}, \& {Wang}}]{Zhang_etal09}
{Zhang}, W., {MacFadyen}, A., \& {Wang}, P. 2009, \apj, 692, L40

\end{thebibliography}

\end{document}